\documentclass[preprint,prd,amsmath,amssymb,floatfix,nofootinbib]{revtex4}
\usepackage[colorlinks=true,urlcolor=black]{hyperref}

\hypersetup{backref = true,pagebackref = true,hyperindex = true, colorlinks = true,
  breaklinks = true, urlcolor = blue, linkcolor = blue, bookmarks = true,
  bookmarksopen = true, citecolor=red}

\usepackage{bm}
\usepackage{color}
\usepackage{dcolumn}
\usepackage{graphics}
\usepackage{graphicx}
\usepackage{subfigure}
\usepackage{slashed}
\usepackage{placeins}

\usepackage{pdfpages}
\usepackage{eso-pic}
\usepackage{everyshi}
\usepackage{pdflscape}

\newcommand{\bea}{\begin{eqnarray}}
\newcommand{\eea}{\end{eqnarray}}
\newcommand{\be}{\begin{equation}}
\newcommand{\ee}{\end{equation}}

\newcommand{\ar}{a_s}

\begin{document}

\title{Heavy quark contributions in Bjorken sum rule with analytic coupling
}
\author{I.R. Gabdrakhmanov$^{1}$, N.A Gramotkov$^{1,2}$, A.V.~Kotikov$^{1}$,  O.V.~Teryaev$^{1}$, D.A. Volkova$^{1,3}$  and I.A.~Zemlyakov$^{4}$}
\affiliation{
  $^1$Bogoliubov Laboratory of Theoretical Physics,
  Joint Institute for Nuclear Research, 141980 Dubna, Russia;\\
$^2$Moscow State University, 119991, Moscow, Russia\\
  $^3$Dubna State University,
  141980 Dubna, Moscow Region, Russia;\\
  $^4$Department of Physics, Universidad Tecnica Federico Santa Maria,\\
Avenida Espana 1680, Valparaiso, Chile}


\begin{abstract}

We consider heavy quark contributions to the polarized Bjorken sum rule.
We found good agreement between the experimental data
and the predictions of analytic QCD.
To satisfy the limit of photoproduction, we use
new representation of the perturbative part of the polarized Bjorken sum rule, proposed recently.

\end{abstract}

\maketitle

\section{Introduction}

Experimental data for the polarized Bjorken sum rule (BSR) $\Gamma^{p-n}_1(Q^2)$ \cite{Bjorken:1966jh}
are now available for a fairly wide range of spacelike momenta squared $Q^2$: 0.021 GeV$^2\leq Q^2 <$5 GeV$^2$
(see \cite{Deur:2021klh,Gabdrakhmanov:2024bje}) and references therein), making the BSR an important
observable for QCD studies at low $Q^2$ \cite{Deur:2018roz,Kuhn:2008sy}.

In the last thirty years, an extension of the QCD coupling constant {\it (couplant)} without the Landau singularity for low $Q^2$,
  called analytic perturbation theory (APT) \cite{ShS,BMS1}, has been developed.
  APT has already been applied to compare theoretical expressions and experimental data of BSR
  \cite{Gabdrakhmanov:2024bje,Pasechnik:2008th,Khandramai:2011zd,Ayala:2017uzx,Ayala:2018ulm,Gabdrakhmanov:2023rjt} (see also other recent
  BSR studies in \cite{Kotlorz:2018bxp,Ayala:2023wpy}).

  In this paper we apply the results for the heavy quark (HQ) contributions to the BSR calculated at the two-loop level
  in Ref. \cite{Blumlein:2016xcy}.
  Our study is carried out in the APT framework, and we also show a possibility of applying the HQ contributions to
  the photoproduction limit.

\section{Bjorken sum rule}

The polarized BSR is defined as the difference between the proton and neutron polarized structure functions,
integrated over the entire interval $x$:
\be
\Gamma_1^{p-n}(Q^2)=\int_0^1 \, dx\, \bigl[g_1^{p}(x,Q^2)-g_1^{n}(x,Q^2)\bigr].
\label{Gpn} 
\ee

Theoretically, since we plan to consider here in particular very low $Q^2$ values, the quantity $\Gamma_1^{p-n}(Q^2)$
can be written in the OPE form with the so-called "massive" twist-four representation (see \cite{Teryaev:2013qba,Gabdrakhmanov:2017dvg}):
\be
\Gamma_1^{p-n}(Q^2)=
\frac{g_A}{6} \, \bigl(1-D_{\rm BS}(Q^2)\bigr) +\frac{\hat{\mu}_4 M^2}{Q^{2}+M^2} \, ,
\label{Gpn.mOPE} 
\ee
where $g_A$=1.2762 $\pm$ 0.0005 \cite{PDG20} is the axial charge of the nucleon,
$(1-D_{BS}(Q^2))$ is the contribution of the leading twist (or twist-two), and the values of $\hat{\mu}_4$ and $M^2$ of the
twist-four term are free parameters that must be determined from experimental data.

Up to the $k$-th order of perturbation theory (PT), the twist-two part has the form
\be
D^{(1)}_{\rm BS}(Q^2)=\frac{4}{\beta_0} \, a^{(1)}_s,~~D^{(k\geq2)}_{\rm BS}(Q^2)=\frac{4}{\beta_0} \, a^{(k)}_s\left(1+\sum_{m=1}^{k-1} d_m \bigl(a^{(k)}_s\bigr)^m
\right)\,,
\label{DBS} 
\ee
where $d_1$, $d_2$ and $d_3$ are eactly known (see, for example, \cite{Chen:2006tw}).
The exact $d_4$ value is not known, but it was  estimated  in Ref. \cite{Ayala:2022mgz}.


{\bf 1.} Following \cite{Cvetic:2006mk}, we introduce and use here the derivatives (in the $k$-order of PT)
\be
\tilde{a}^{(k)}_{n+1}(Q^2)=\frac{(-1)^n}{n!} \, \frac{d^n a^{(k)}_s(Q^2)}{(dL)^n},~~a^{(k)}_s(Q^2)=\frac{\beta_0 \alpha^{(k)}_s(Q^2)}{4\pi}=\beta_0\,\overline{a}^{(k)}_s(Q^2),
\label{tan+1}
\ee
which play a key role for the construction of analytic QCD. 
Hereafter $\beta_0$ is the first coefficient of the QCD $\beta$-function:
\be
\beta(\overline{a}^{(k)}_s)=-{\left(\overline{a}^{(k)}_s\right)}^2 \bigl(\beta_0 + \sum_{i=1}^k \beta_i {\left(\overline{a}^{(k)}_s\right)}^i\bigr),
\label{bQCD}
\ee
where $\beta_i$ are known up to $k=4$ \cite{Baikov:2008jh}.

The series of derivatives of $\tilde{a}_{n}(Q^2)$ can be used instead of the series of $\ar$-powers. Indeed, although each derivative reduces the $\ar$ power,
on the other hand it produces an additional $\beta$-function and hence an additional $\ar^2$ factor.
By definition (\ref{tan+1}), in the leading order (LO) the expressions for $\tilde{a}_{n}(Q^2)$ and $\ar^{n}$ coincide exactly.
Beyond LO, there is a one-to-one correspondence between $\tilde{a}_{n}(Q^2)$ and $\ar^{n}$, established in \cite{Cvetic:2006mk,Cvetic:2010di} and
extended to the fractional case in \cite{GCAK}.


{\bf 2.} Converting the couplant powers into its derivatives, we have
\be
D^{(1)}_{\rm BS}(Q^2)=\frac{4}{\beta_0} \, \tilde{a}^{(1)}_1,~~D^{(k\geq2)}_{\rm BS}(Q^2)=
\frac{4}{\beta_0} \, \left(\tilde{a}^{(k)}_{1}+\sum_{m=2}^k\tilde{d}_{m-1}\tilde{a}^{(k)}_{m}
\right),
\label{DBS.1} 
\ee
where
\bea
&&\tilde{d}_1=d_1,~~\tilde{d}_2=d_2-b_1d_1,~~\tilde{d}_3=d_3-\frac{5}{2}b_1d_2-\bigl(b_2-\frac{5}{2}b^2_1\bigr)\,d_1,\nonumber \\
&&\tilde{d}_4=d_4-\frac{13}{3}b_1d_3 -\bigl(3b_2-\frac{28}{3}b^2_1\bigr)\,d_2-\bigl(b_3-\frac{22}{3}b_1b_2+\frac{28}{3}b^3_1\bigr)\,d_1
\label{tdi} 
\eea
and $b_i=\beta_i/\beta_0^{i+1}$.

For the case of 3 active quark flavors ($f=3$), which is approved in this work, we have
\footnote{
  The coefficients $\beta_i$ $(i\geq 0)$ of the $\beta$ function (\ref{bQCD}) and hence the couplant $\alpha_s(Q^2)$ itself
  depend on the number $f$, and each new quark enters/leaves the game at a certain threshold $Q^2_f$ according to \cite{Chetyrkin:2005ia}.
  The corresponding parameters $\Lambda^{(f)}$ in N$^i$LO PT can be found in \cite{Chen:2021tjz}.}
\bea
&&d_1=1.59,~~d_2=3.99,~~d_3=15.42~~d_4=63.76, \nonumber \\ 
&&\tilde{d}_1=1.59,~~\tilde{d}_2=2.73,
~~\tilde{d}_3=8.61,~~\tilde{d}_4=21.52 \, ,
\label{td123} 
\eea
i.e. the coefficients in the derivative series are slightly smaller.

{\bf 3.} {\it HQ contribution}
was calculated in \cite{Blumlein:2016xcy} only at the next-to-leading (NLO) order, that
leads to the following replacement for $\tilde{d}_1=d_1$:
\be
d_1 \to d_1 - \sum_{i=c,b,t}\,C_1(\xi_i),~~\tilde{d}_1 \to \tilde{d}_1 - \sum_{i=c,b,t}\,C_1(\xi_i)\,,
    \label{Gpn.MA.HQ} 
\ee
where
\be
\xi_i=\frac{Q^2}{m^2_i}~~(i=c,b,t)
\label{xi} 
\ee
and $m_c=1.27$ GeV, $m_b=4.18$ GeV and $m_t=172.76$ GeV (see \cite{PDG20}).

$C_1(\xi)$ has the followng form
\bea
C_1(\xi)&=&\frac{8}{3\beta_0} \, \biggl\{\frac{6\xi^2+2735\xi+11724}{5040\xi}- \frac{3\xi^3+106\xi^2+1054\xi+4812}{2520\xi}\,L(\xi)\nonumber\\ 
&-&\frac{5}{3\xi(\xi+4)}\,L^2(\xi)+\frac{3\xi^2+112\xi+1260}{5040}\,\ln(\xi)\biggl\}\,,
\label{C1} 
\eea
where
\be
L(\xi)=\frac{1}{2\delta}\,\ln\left(\frac{1+\delta}{1-\delta}\right),~~\delta^2=\frac{\xi}{4+\xi}\,.
\label{L} 
\ee

At large $Q^2$ values, we have 
\be
C_1(\xi)=\frac{2}{3\beta_0} \, \biggl\{\frac{1}{2}-\frac{5}{12\xi^2}\,\ln^2(\xi)-\frac{4}{3\xi}\,\ln(\xi)+\frac{17}{9\xi}
+O\left(\frac{\ln(\xi)}{\xi^2}\right)
\biggl\}\,,
\label{C1approxL} 
\ee
i.e. there is the HQ decoupling here. 

At low $Q^2$ values, there is the approximation
\be
C_1(\xi)=\frac{2}{3\beta_0} \, \biggl\{\left(1+\frac{4\xi}{45}+\frac{\xi^2}{420}\right)\,\ln(\xi)- \frac{58\xi}{225}-\frac{1933\xi^2}{176400}
+O\bigl(\xi^3\bigr)
\biggl\}\,,
\label{C1approx} 
\ee
which shows that the contribution increases as $\ln Q^2$ at $Q^2 \to 0$ and there is no separation within PT.
\footnote{We are grateful to J. Blumlein for explaining the problem to us.}


{\bf 4.}
In APT, the results (\ref{Gpn.mOPE}) become as follows
\be
\Gamma_{\rm{A},1}^{p-n}(Q^2)=
\frac{g_A}{6} \, \bigl(1-D_{\rm{A,BS}}(Q^2)\bigr) +\frac{\hat{\mu}_{\rm{A},4}M^2}{Q^{2}+M^2},~~
\label{Gpn.MA} 
\ee
where the perturbative part $D_{\rm{BS,A}}(Q^2)$ takes the same form as (\ref{DBS.1}), but with the analytic couplant $\tilde{A}^{(k)}_{\nu}$
(the corresponding expressions for $\tilde{A}^{(k)}_{\nu}$ can be found \cite{Kotikov:2022sos,Kotikov:2023meh})
\be
D^{(1)}_{\rm A,BS}(Q^2)=\frac{4}{\beta_0} \, A^{(1)},~~
D^{k\geq2}_{\rm{A,BS}}(Q^2) =\frac{4}{\beta_0} \, \Bigl(A^{(1)}
+ \sum_{m=2}^{k} \, \tilde{d}_{m-1} \, \tilde{A}^{(k)}_{\nu=m} \Bigr)\,.
\label{DBS.ma} 
\ee

\section{Results}

\begin{table}[t]
\begin{center}
\begin{tabular}{|c|c|c|c|}
\hline
& $M^2$ for $Q^2 \leq 5$ GeV$^2$ & $\hat{\mu}_{\rm{MA},4}$  for $Q^2 \leq 5$ GeV$^2$& $\chi^2/({\rm d.o.f.})$ for $Q^2 \leq 5$ GeV$^2$ \\
& (for $Q^2 \leq 0.6$ GeV$^2$) & (for $Q^2 \leq 0.6$ GeV$^2$) & (for $Q^2 \leq 0.6$ GeV$^2$) \\
 \hline
 LO & 0.472 $\pm$ 0.035 & -0.212 $\pm$ 0.006 & 0.667  \\
 & (1.631 $\pm$ 0.301) & (-0.166 $\pm$ 0.001) & (0.789)  \\
 \hline
 NLO & 0.392 $\pm$ 0.036 & -0.196 $\pm$ 0.008 & 0.759  \\
& (1.740 $\pm$ 0.389) & (-0.143 $\pm$ 0.002) & (0.742)  \\
 \hline
 N$^2$LO & 0.374 $\pm$ 0.036 & -0.198$\pm$ 0.008 & 0.781  \\
 & (1.574 $\pm$ 0.319) & (-0.144 $\pm$ 0.002) & (0.714)  \\
 \hline
 N$^3$LO & 0.372 $\pm$ 0.036 & -0.200 $\pm$ 0.009 & 0.789  \\
  & (1.588 $\pm$ 0.327) & (-0.145 $\pm$ 0.002) & (0.733)  \\
   \hline
N$^4$LO & 0.374 $\pm$ 0.036 & -0.199 $\pm$ 0.009 & 0.789  \\
 & (1.630 $\pm$ 0.344) & (-0.144 $\pm$ 0.002) & (0.739)  \\
 \hline
\end{tabular}
\end{center}
\caption{%
  The values of the fit parameters in (\ref{Gpn.MA}).
}
\label{Tab:BSR}
\end{table}

\begin{figure}[!htb]
\centering
\includegraphics[width=0.98\textwidth]{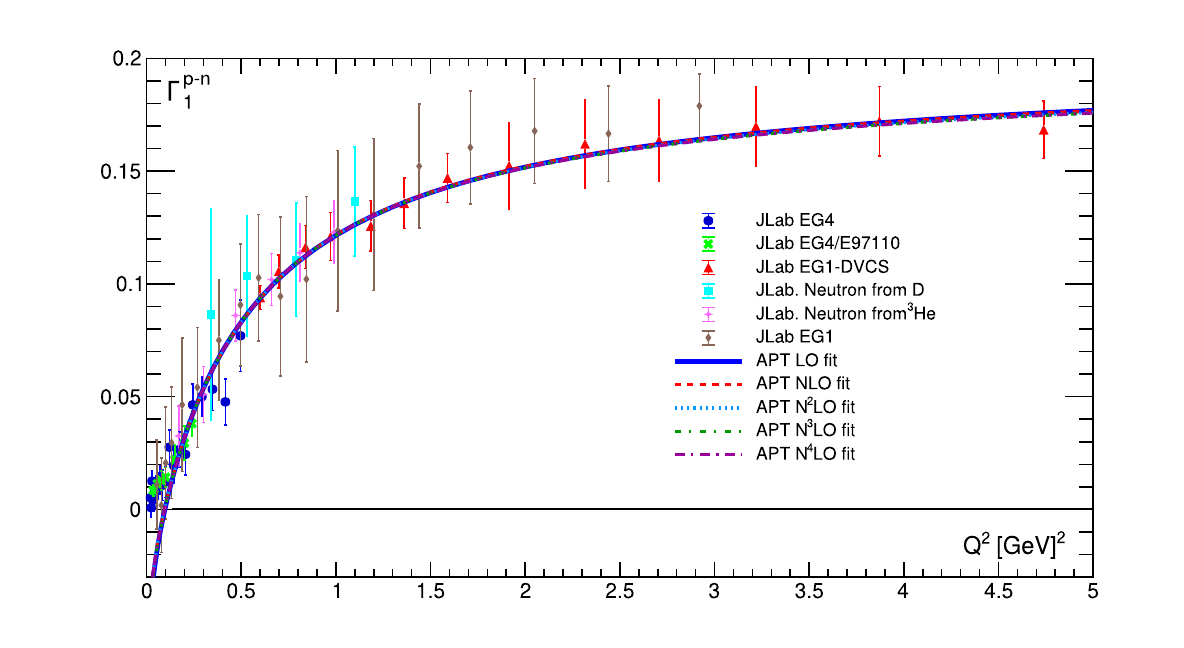}
\caption{
  \label{fig:APT}
  The results for $\Gamma_1^{p-n}(Q^2)$ (\ref{Gpn.MA}) in the first  four
  orders of APT.
    }
\end{figure}

Since the regular PT is not applicable for low-$Q^2$ BSRs
(see \cite{Gabdrakhmanov:2024bje,Pasechnik:2008th,Khandramai:2011zd,Ayala:2017uzx,Ayala:2018ulm,Gabdrakhmanov:2023rjt}), we consider only APT here.
The fitting results  of experimental data obtained only with
statistical uncertainties 
are presented in Table \ref{Tab:BSR} and shown in Figs. 1 and 2.
Following \cite{Gabdrakhmanov:2024bje,Gabdrakhmanov:2023rjt}, we consider fits to the full data set as well as to the $Q^2~<$0.6 GeV$^2$ data.

Our results obtained for different APT orders are almost equivalent: the corresponding curves become
indistinguishable when $Q^2$ approaches 0, and differ slightly elsewhere. As can be seen in Figa. 1 and 2,
the quality of the fit is quite good, as demonstrated by the values of the corresponding
$\chi^2/({\rm d.o.f.})$ (see Table \ref{Tab:BSR}).

As in the case without heavy quarks, considered in \cite{Gabdrakhmanov:2024bje,Gabdrakhmanov:2023rjt},
the situation is more complicated, however, as shown in Fig. 2.
The curves obtained by the fits take negative values when we go to very low values of $Q^2$: $Q^2 <$0.02 GeV$^2$.
The reason for this phenomenon can be shown by considering photoproduction in the APT framework, which is the topic of the next subsection.

\begin{figure}[t]
\centering
\includegraphics[width=0.98\textwidth]{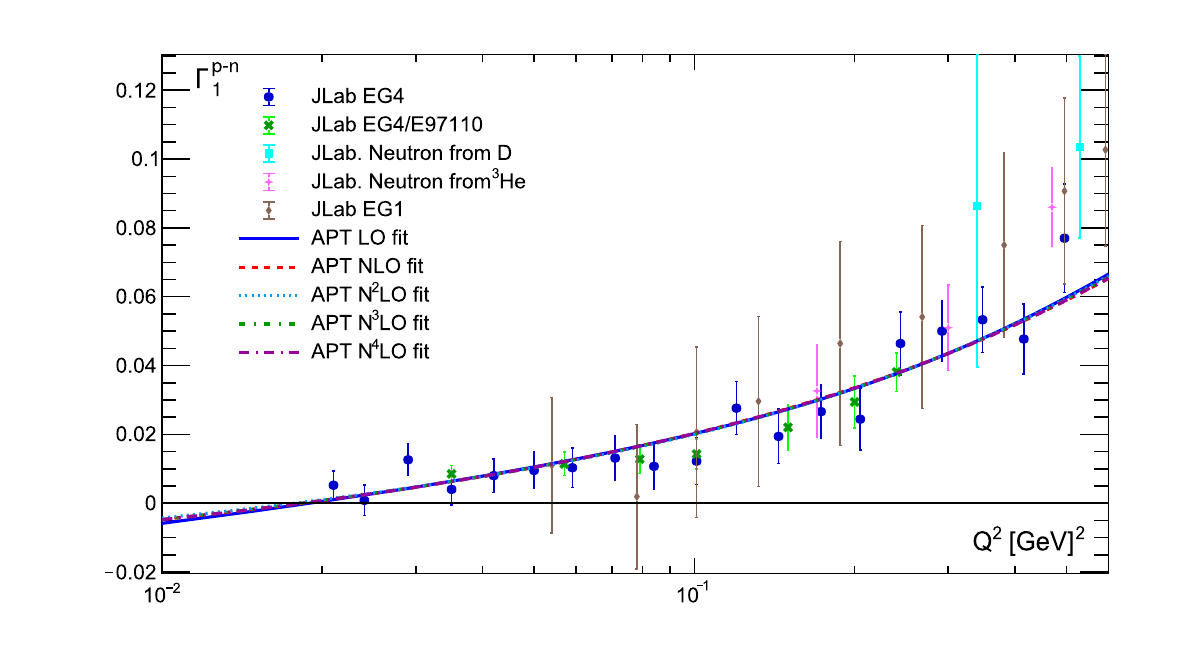}
\caption{
  \label{fig:low}
As in Fig 1 but
with $Q^2 <$0.6 GeV$^2$
}
\end{figure}

{\bf 1.} {\it Photoproduction.}
To obtain the correct limit of $\Gamma_{\rm{A},1}^{p-n}(Q^2\to 0)\to 0$, coming from the finitness of photoproduction cross-section,
a new form for $\Gamma_{\rm{MA},1}^{p-n}(Q^2)$ was proposed in \cite{Gabdrakhmanov:2024bje}:
\be
\Gamma_{\rm{A},1}^{p-n}(Q^2)=
\, \frac{g_{A}}{6} \, \bigl(1-D_{\rm{A,BS}}(Q^2) \cdot \frac{Q^2}{Q^2+M^2}\bigr) +
\frac{\hat{\mu}_{\rm{A},4}M^2}{Q^{2}+M^2}+\frac{\hat{\mu}_{\rm{A},6}M^4}{(Q^{2}+M^2)^2},~~
\label{Gpn.MAn} 
\ee
where we added a ``massive'' twist-six term and introduced a factor $Q^2/(Q^2+M^2)$ to modify the twist-two part.

\begin{table}[t]
\begin{center}
\begin{tabular}{|c|c|c|}
\hline
& $M^2$ for $Q^2 \leq 5$ GeV$^2$& $\chi^2/({\rm d.o.f.})$ for $Q^2 \leq 5$ GeV$^2$ \\
& (for $Q^2 \leq 0.6$ GeV$^2$)  & (for $Q^2 \leq 0.6$ GeV$^2$) \\
 \hline
 LO & 0.383 $\pm$ 0.014 (0.576 $\pm$ 0.046) & 0.572 (0.575) \\
 \hline
 NLO & 0.319 $\pm$ 0.014  (0.411 $\pm$ 0.035) & 1.104 (0.630) \\
 \hline
 N$^2$LO & 0.308 $\pm$ 0.014 (0.400 $\pm$ 0.034) & 1.159 (0.621) \\
 \hline
 N$^3$LO & 0.309 $\pm$ 0.014  (0.411 $\pm$ 0.035) & 1.181 (0.618) \\
   \hline
N$^4$LO & 0.311 $\pm$ 0.014  (0.412 $\pm$ 0.035) & 1.174  (0.621) \\
 \hline
\end{tabular}
\end{center}
\caption{%
  The values of the fit parameters in (\ref{Gpn.MAn}).
}
\label{Tab:BSR1}
\end{table}

The finitness of cross-section in the real photon limit  leads now to (see \cite{Gabdrakhmanov:2024bje})
\bea
\hat{\mu}_{\rm{A},6} =-G\,M^2+\frac{5g_{A}}{54}= -G\,M^2+0.1182,\nonumber \\
\hat{\mu}_{\rm{A},4} = -\frac{g_{A}}{6} -\hat{\mu}_{\rm{A},6}= G\,M^2-\frac{7g_{A}}{27}=  G\,M^2-0.3309\,,
\label{Gpn.MAnQ0.4} 
\eea
where \cite{Soffer:1992ck,Pasechnik:2010fg}
\be
G=\frac{\mu^2_n-(\mu_p-1)^2}{8M_p^2}=0.0631\,,
  \label{GDH} 
\ee
which is small and, thus,
$\hat{\mu}_{\rm{MA},4}<0$ and $\hat{\mu}_{\rm{MA},4} 
>0$.

The results of fitting the theoretical predictions based on(\ref{Gpn.MAn}) with $\hat{\mu}_{\rm{MA},4}$ and $\hat{\mu}_{\rm{MA},6}$ from
(\ref{Gpn.MAnQ0.4}) are presented in Table \ref{Tab:BSR1} and in Figs. 3 and 4.

As can be seen from Table \ref{Tab:BSR1}, the results obtained for $M^2$ differ when using the full data set
and the limited one with $Q^2<$0.6 GeV$^2$.
However, the difference is much smaller than in Table \ref{Tab:BSR}.
Moreover, the results are very similar to those obtained in \cite{Gabdrakhmanov:2024bje,Gabdrakhmanov:2023rjt}, since the increase in
$C_1(\xi)$ as $Q^2 \to 0$ is compensated by the decrease in $\tilde{A}^{(k)}_{\nu=2}(Q^2 \to 0)$.

We also see similarities between the results shown in Figs. 2 and 4. The difference appears only for small values of $Q^2$.
Fig. 4 also shows that the results of fitting the full set of experimental data agree better with the data at $Q^2\geq 0.55$GeV$^2$,
as they should since these data are included in the analysis.

\begin{figure}[t]
\centering
\includegraphics[width=0.98\textwidth]{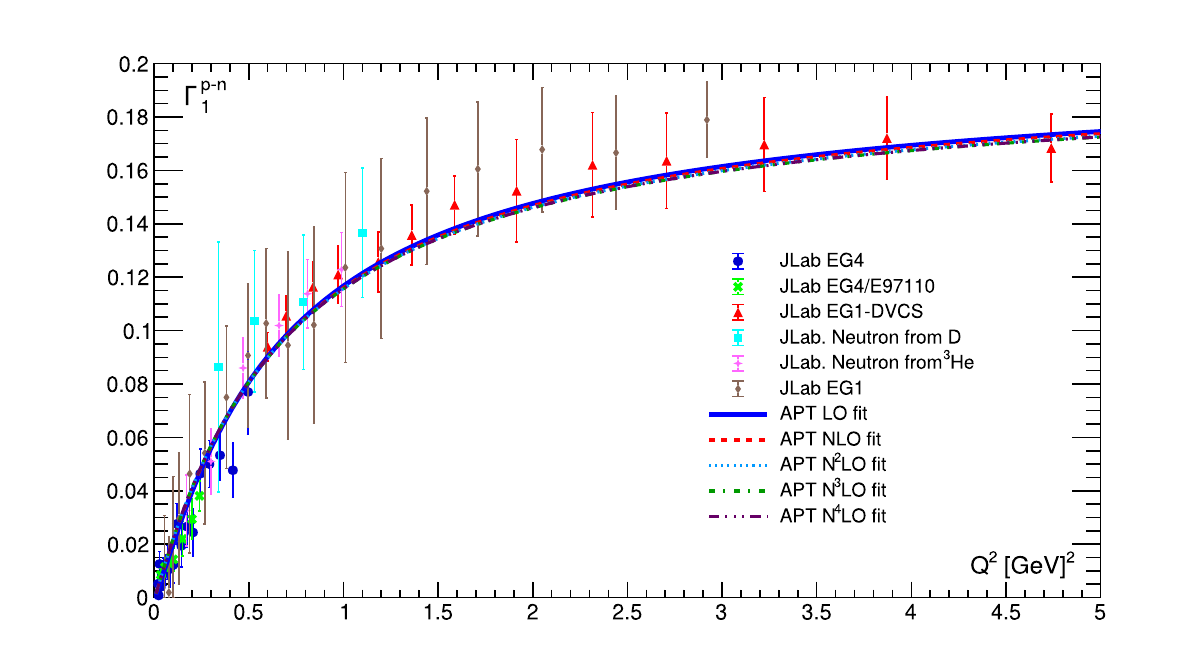}
\caption{
  \label{fig:PT1}
  The results for $\Gamma_1^{p-n}(Q^2)$ (\ref{Gpn.MAn})
  in the first  four  orders of APT. 
}
\end{figure}

\begin{figure}[t]
\centering
\includegraphics[width=0.98\textwidth]{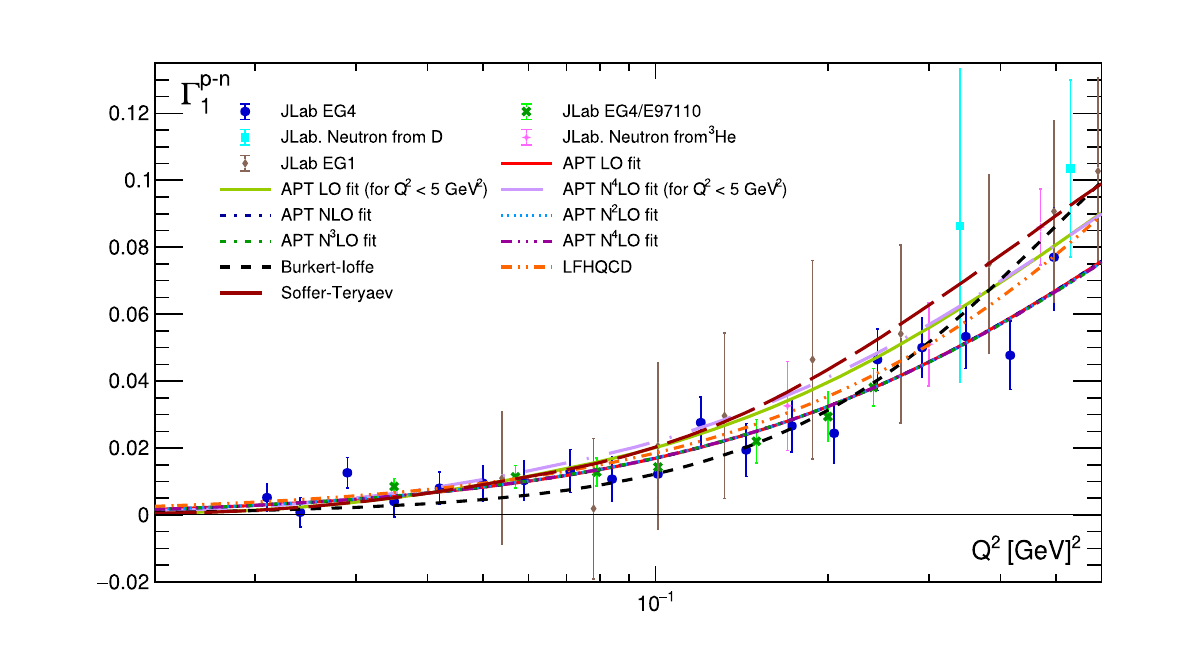}
\caption{
  \label{fig:low1}
  As in Fig. \ref{fig:PT1} but for $Q^2<$0.6 GeV$^2$
}
\end{figure}

\section{Conclusions}

We examined the Bjorken sum rule $\Gamma_{1}^{p-n}(Q^2)$ taking into account the contribution of heavy quarks within
the APT framework and found good agreement with experimental data.

By investigating the low $Q^2$ behavior, we found, as in previous studies without heavy quarks, that there is a
discrepancy between the results obtained in the fits and the photoproduction. Indeed, the results of the fits
extended to low $Q^2$ lead to negative values for the Bjorken sum rule $\Gamma_{\rm{MA},1}^{p-n}(Q^2)$:
$\Gamma_{\rm{MA},1}^{p-n}(Q^2\to 0) <0$, which contradicts the finiteness of the cross section in the real photon limit,
leading to $\Gamma_{\rm{MA},1}^{p-n}(Q^2\to 0) =0$.

To solve the problem, we used a low $Q^2$ modification of the OPE formula for $\Gamma_{\rm{MA},1}^{p-n}(Q^2)$ introduced in \cite{Gabdrakhmanov:2024bje}.
Using it, we found good agreement with full sets of experimental data for the Bjorken sum rule
$\Gamma_{\rm{MA},1}^{p-n}(Q^2)$, as well as with its limit $Q^2 \to 0$, i.e. with photoproduction.
Moreover, the results are very close to those obtained in \cite{Gabdrakhmanov:2024bje,Gabdrakhmanov:2023rjt}, as well as to the predictions
of phenomenological models \cite{Soffer:1992ck,Pasechnik:2010fg,Burkert:1992tg,Brodsky:2014yha}.


{\bf Acknowledgments.}~Authors are grateful to Alexandre P. Deur for
information about new experimental data in Ref. \cite{Deur:2021klh} and Johannes Blumlein for initiating the consideration of the contribution of heavy quarks.
Authors thank Konstantin Chetyrkina and Andrei Kataev for careful discussions.
This work was supported in part by the Foundation for the Advancement of Theoretical
Physics and Mathematics “BASIS”.
One of us (I.A.Z.)
is supported by the Directorate of
Postgraduate Studies of the Technical University of Federico Santa Maria.



\end{document}